\documentclass[sigconf]{acmart}

\usepackage{booktabs} 
\usepackage{graphicx}
\usepackage{caption}
\copyrightyear{2019} 
\acmYear{2019} 
\setcopyright{acmlicensed}
\acmConference[WebSci '19 Companion]{11th ACM Conference on Web Science Companion}{June 30-July 3, 2019}{Boston, MA, USA}
\acmBooktitle{11th ACM Conference on Web Science Companion (WebSci '19 Companion), June 30-July 3, 2019, Boston, MA, USA}
\acmPrice{15.00}
\acmDOI{10.1145/3328413.3328416}
\acmISBN{978-1-4503-6174-3/19/06}

\begin{document}
\title{On Bias in Social Reviews of University Courses}

\author{Taha Hassan}
\affiliation{%
  \institution{Computer Science Department, Virginia Tech}
  \city{Blacksburg}
  \state{VA}
}
\email{taha@vt.edu}


\renewcommand{\shortauthors}{T. Hassan}

\begin{abstract}
University course ranking forums are a popular means of disseminating information about satisfaction with the quality of course content and instruction, especially with undergraduate students. A variety of policy decisions by university administrators, instructional designers and teaching staff affect how students perceive the efficacy of pedagogies employed in a given course, in class and online. While there is a large body of research on qualitative driving factors behind the use of academic rating sites, there is little investigation of the (potential) implicit student bias on said forums towards desirable course outcomes at the institution level. To that end, we examine the connection between course outcomes (student-reported GPA) and the overall ranking of the primary course instructor, as well as rating disparity by nature of course outcomes, for several hundred courses taught at Virginia Tech based on data collected from a popular academic rating forum. We also replicate our analysis for several public universities across the US. Our experiments indicate that there is a discernible albeit complex bias towards course outcomes in the professor ratings registered by students.

\end{abstract}




\maketitle
\section{Introduction}
Online forums for rating university course instructors, like \textit{RateMyProfessors} \cite{rmphomepage} and \textit{Koofers} \cite{koofershomepage} have inspired considerable research attention over the years \cite{chang2014koofers} \cite{brown2009rating} \cite{legg2012ratemyprofessors} \cite{kindred2005he}. An aggregate of content sharing, networking and recommendation services, these forums cater to a number of their contributors' needs, including but not limited to information seeking, gratification, and convenience \cite{kindred2005he} . Prior studies have identified several broad themes in the student feedback sampled from these forums, including teacher personality, aptitude and preparation, ease of access to help and feedback from course staff, and percevied practicality of the course rubric \cite{hartman2013ratemyprofessors}. There is however, lesser attention devoted in this literature, to institution-level correlates of student perception. Empirical investigations of the potential sources of student bias in said perception are often divided in their conclusions, because of limitations of sample sizes or meta-variable space \cite{marsh1984} \cite{centragradestudy} \cite{feldman2007review}. Assessing its reliability at scale can lend insights to instructional designers, department administrators and instructors alike on the limitations of existing pedagogies. It can also potentially extend the utility of university-managed end-of-semester course evaluations, and help improve the usability, relevance, accessibility and trustworthiness \cite{legg2012ratemyprofessors}\cite{hassan2019trust} of its host forums.
\begin{figure}
\includegraphics[width=0.5\textwidth]{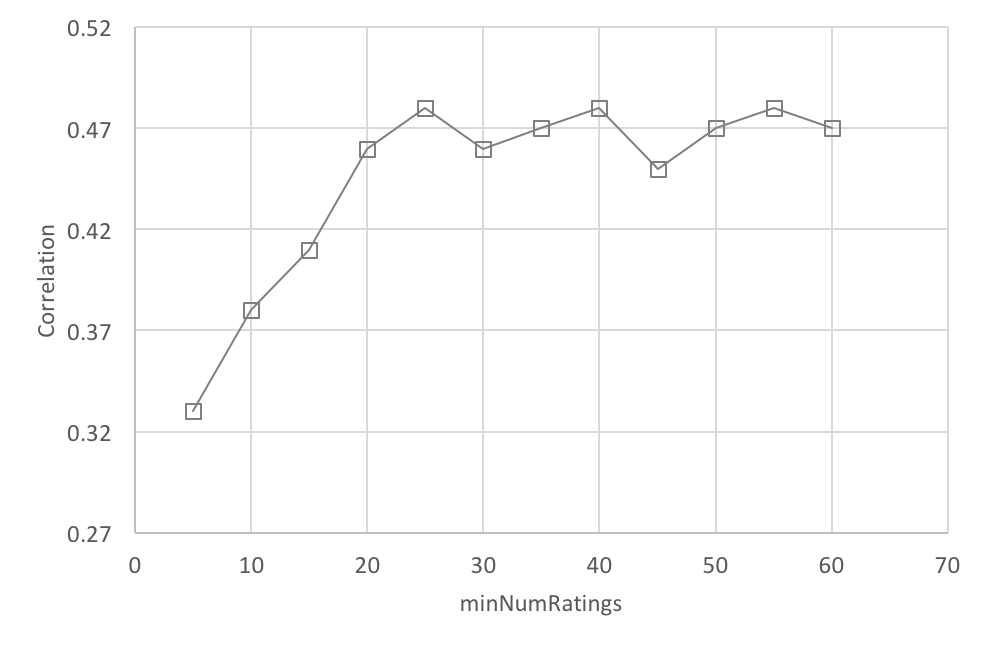}
\caption{Correlation between the average student-reported GPA for a course and overall instructor rating, as a function of the minimum number of ratings considered}
\label{fig:gap}
\end{figure} 




\begin{table}[]
\caption{Key counts for courses in the historical ratings dataset}
\centering
\label{tab_key_counts}
\begin{tabular}{p{2.0cm} p{1.2cm} p{1.8cm} p{1.2cm}}
\toprule
\bfseries Institution & \bfseries Courses & \bfseries Departments & \bfseries Ratings \\
\midrule
MSU & 888 & 110 & 67086\\
VT & 597 & 80 & 42503\\
CAL-POLY & 578 & 67 & 36900\\
CSU & 430 & 74 & 20506\\
UGA & 367 & 99 & 22992\\
UNC-CH & 84 & 34 & 2180\\
GMU & 75 & 37 & 1866\\
JMU & 72 & 33 & 2158\\
UKY & 69 & 30 & 1537\\
NCSU & 61 & 29 & 1923\\
-----------------\\
VCU & 39 & 20 & 939\\
UVA & 24 & 14 & 581\\
ASU & 13 & 10 & 236\\
\bottomrule 
\end{tabular}
\end{table}

\section{Approach}
We pursue preliminary evidence of what appears to be a modest to strong relative connection between aggregate course outcomes and student perception of the course instructor. Figure \ref{fig:gap} visualizes the correlation between these two for 478 courses taught at Virginia Tech. As we increase the minimum number of ratings per course considered towards the correlation, this correlation increases and achieves a steady average value of about 0.47 beyond 20 ratings per course. We explore this further by examining the disparity of instructor ranking between high, medium and low GPA student groups, as well as regressing student approval against course outcomes and perceived difficulty of various course instruments.

\begin{table*}[ht]
\caption{Hypothesis-testing the relationship b/w course outcomes and instructor rankings}
\centering
\label{tab_hyp}
\begin{tabular}{p{2.8cm} p{1.8cm} p{2.8cm} p{1.0cm} p{1.4cm} p{1.4cm} p{1.4cm} p{1.4cm}}
\toprule
\bfseries Institution & \bfseries $Corr, p$ & $F (df_1, df_2), p$ & $F_{crit}$ & $\mu_{ov}, N_{ov}$ & $\mu_{hi}, N_{h}$ & $\mu_{med}, N_{m}$ & $\mu_{low}, N_{l}$\\
\midrule
MSU & 0.24, 3.1e-13* & 14.7 (2, 885), 5e-7$^{\dagger}$ & 3.0 & 3.82, 888 & 3.88, 64 & 3.85, 741 & 3.46, 83\\
VT & 0.33, 4e-17* & 15.3 (2, 594), 3.1e-7$^{\dagger}$ & 3.01 & 3.83, 597 & 4.15, 30 & 3.86, 492 & 3.52, 75\\
CAL-POLY & 0.23, 2.8e-8* & 16.7 (2, 575), 8.8e-8$^{\dagger}$ & 3.01 & 3.89, 578 & 4.39, 17 & 3.92, 397 & 3.78, 164\\
CSU & 0.27, 6.8e-9* & 15.6 (2, 427), 2.7e-7$^{\dagger}$ & 3.01 & 3.72, 430 & 4.14, 16 & 3.79, 291 & 3.5, 123\\
UGA & 0.41, 3e-16* & 19.4 (2, 364), 9e-9$^{\dagger}$ & 3.02 & 3.84, 367 & 4.32, 43 & 3.87, 302 & 3.59, 22\\
UNC-CH & 0.48, 4.8e-6* & 6.6 (2, 81), 2e-3$^{\dagger}$ & 3.1 & 3.69, 84 & 4.65, 2 & 3.73, 68 & 3.35, 14\\
GMU & 0.5, 4.5e-6* & 8 (1, 73), 6e-3$^{\dagger}$ & 3.12 & 3.78, 75 & - & 3.95, 42 & 3.57, 33\\
JMU & 0.46, 4.2e-5* & 3.8 (2, 69), 2e-2$^{\dagger}$ & 3.13 & 3.74, 72 & 4.05, 1 & 3.82, 45 & 3.58, 26\\
UKY & -0.09, 0.49 & 0.13 (2, 66), 0.71 & 3.14 & 3.54, 69 & 3.17, 1 & 3.52, 41 & 3.6, 27\\
NCSU & 0.39, 2e-3* & 4.3 (2, 58), 1.7e-2$^{\dagger}$ & 3.15 & 3.72, 61 & 4.43, 3 & 3.75, 39 & 3.53, 19\\
-----------------\\
VCU & 0.1, 0.53 & 3.5 (1, 37), 0.06 & 3.26 & 3.82, 39 & - & 4.04, 9 & 3.75, 30\\
UVA & 0.14, 0.52 & 0.62 (1, 22), 0.43 & 3.44 & 3.73, 24 & 4.62, 1 & 3.7, 23 & -\\
ASU & 0.74, 3e-3* & 0.07 (1, 11), 0.78 & 3.98 & 3.91, 13 & - & 4.14, 7 & 3.64, 6\\
-----------------\\
\multicolumn{3}{l}{* stat. significant, $\alpha=0.05$} & & & & \\
\multicolumn{3}{l}{$\dagger$ stat. significant, $F>F_{crit}, \alpha=0.05$} & & & & \\
\bottomrule 
\end{tabular}
\end{table*}

\begin{table}[ht]
\caption{Regression analysis: overall professor rating, function of student GPA (\textbf{X1}), and perceived ease of course instruments (\textbf{X2} - \textbf{X5})}
\centering
\label{tab_reg}
\begin{tabular}{p{2.2cm} p{1.2cm} p{1.2cm} p{0.8cm} p{0.8cm}}
\toprule
 & \textbf{coef} & \textbf{std. error} & $t$ & $p$\\
\midrule
\textbf{intercept} & 2.6 & 0.318 & 8.3 & 0.00*\\
\textbf{X1}: GPA & 0.5 & 0.085 & 6.0 & 0.00*\\
\textbf{X2}: exams & -0.03 & 0.041 & -0.8 & 0.398\\
\textbf{X3}: quizzes & -0.08 & 0.037 & -2.3 & 0.02*\\
\textbf{X4}: projects & -0.04 & 0.027 & -1.7 & 0.08\\
\textbf{X5}: homework & -6e-3 & 0.032 & -0.19 & 0.84\\
\bottomrule 
\end{tabular}
\end{table}

\begin{table}[ht]
\caption{Regression analysis (cont.)}
\centering
\label{tab_reg2}
\begin{tabular}{p{3.0cm} p{1.5cm}}
\toprule
stat. & val.\\
\midrule
\textbf{R-squared} & 0.193\\
\textbf{Adj R-squared} & 0.182\\
\textbf{F-statistic} & 17.8\\
\textbf{Prob (F-statistic)} & 8e-16*\\
\textbf{Log-likelihood} & -244.4\\
\textbf{AIC} & 500.8\\
\textbf{BIC} & 524.4\\
\bottomrule 
\end{tabular}
\end{table}

\section{Evaluation}
\subsection{Datasets}
We scraped course metadata from Koofers \cite{koofershomepage}, a popular forum for sharing course content and instructor reviews (see table \ref{tab_key_counts}). For a university, we consider all courses with a minimum of 10 ratings. A course can have a multitude of instructors and offerings. Instructor ratings on Koofers are on a 0 to 5 scale, GPA reports are on a 4.0 scale. We define the minimum acceptable use of the forum (at the institution level) as an excess of 1000 total ratings, with the course count at least twice that of the department count. 

\subsection{Methods}
We significance-test the disparity in professor ratings by GPA groups using one-way ANOVA (F-test, table \ref{tab_hyp}). We also use the Python package \textit{statsmodels.OLS} towards regression analysis of instructor ratings as a function of student GPA, as well as their self-reported ease of course instruments (exams, quizzes, projects and homeworks). 

\subsection{Results}
Table \ref{tab_hyp} lists the correlations between average professor rating and student GPA for all institutions considered, as well as average ratings for each GPA group. The top four institutions with the largest set of ratings (MSU, VT, CAL-POLY, and CSU) register a modest correlation (between 0.2 and 0.35) and large group disparity by GPA. MSU, for instance, reports the highest disparity ($F(2, 885) = 14.7, p=5e-7$) against $F_{crit} = 3$. The next four institutions (UGA, UNC, GMU and JMU) report stronger correlations (between 0.4 and 0.5), with weaker but significant group disparities. It is also instructive to consider the differences between the two groups of courses. The group with minimum acceptable use of the forum almost consistently registers a correlation between outcomes and student perception of the primary instructor, as well as a disparity between ratings of the GPA groups. The group with less than minimum use of the forum generally does not. Overall and department-level sample sizes decline fairly rapidly, especially for the final five institutions in table \ref{tab_hyp}, which we believe, contributes to the larger uncertainty in the corresponding measurements, empty GPA groups and smaller effects sizes if any. 

Table \ref{tab_reg} and table \ref{tab_reg2} report the coefficients, errors and significance tests for regression analysis with average instructor rating as a function of average student-reported GPA \textbf{X1}, average perceived difficulty of exams, quizzes, projects and homeworks (\textbf{X2} through \textbf{X5}). Course outcomes outweigh the perceived ease of course evaluations in their aggregate effect on instructor ratings ($t=6.0, t=-2.3$ for \textbf{X1} and \textbf{X3}, respectively). Difficulty of quizzes appears modestly relevant in ascertaining the overall student satisfaction with the course, with higher difficulty linked to lower student approval (t-statistic is negative). An in-depth analysis of frequent course rubrics and their student approval is left for future work.

\section{Discussion}
Several studies over the last few decades have reported a modest correlation between student evaluations and their grades, albeit for individual students or a course \cite{stumpf_apa} \cite{centragradestudy} \cite{feldman1989association} \cite{feldman2007review}. A review by Feldman \cite{feldman2007review} reports this correlation to be somewhere between +0.1 and +0.3. Our preliminary inquiry at the institution level demonstrates that this correlation matches, and often exceeds said figures. The aforementioned work notes that given the learning acquired by students during the course of a given class or academic term, all of this observed correlation can not necessarily be a result of implicit, time-invariant student bias towards course outcomes. The multi-faceted nature of student perception can affect this connection. This complexity is echoed in a comparative study of in-class assessments, and pre- and post-assessment ratings on RateMyProfessors \cite{legg2012ratemyprofessors}. The study reported how pre-assessment course ratings on instructor clarity were significantly lower than both in-class and post-assessment reviews. However, instructor easiness was reviewed lower in-class relative to online. Our study attempts to initiate a line of large-scale contextual inquiry of these ratings across institutions that can potentially help consolidate these differing interpretations. To that end, we discuss some limitations of our study design, and plans for future research as follows.

\subsection{Limitations} 
We intend to expand our analysis by considering the time order of the instructor ratings in our dataset. While the magnitude of the observed correlation between course outcomes and student rankings is nearly consistent across institutions (with minimum aggregate forum use) we tested for, it is harder to argue about the directionality of this correlation without said data. Another critical deficiency in our approach is the assumption of linearly independent course characteristics. Fixes include dimensionality reduction \cite{scholkopf1997kernel} and modeling the joint reliability of course features using simple Bayesian networks \cite{meila1999accelerated} and are left for future work.

\subsection{Future Work}
Beyond algorithmic improvements, we are working on expanding our dataset to include a larger number of academic institutions, and mine review data to determine whether the following factors affect aggregate student perception of course instructors.

\begin{itemize}
\item university mandate (teaching vs. research, public vs. private),
\item course modality (STEM/non-STEM, undergraduate/graduate, in-class/online)
\item technology use (LMS and third-party apps for course management, testing and assessment)
\item logistics (instructional design training, number of TAs, etc.)
\item forum features - content creation (editorial control, authentication using university credentials)
\item forum features - interaction (accessibility, review search, social tagging)
\item forum features - content management and bias-correction (spam filtering, detection of cyberbullying and defamation)
\end{itemize}

An important step in realizing this contextual inquiry is designing metrics that summarize the observed disparity between effects of course rubric, content quality, interaction fidelity of host forums as well as course outcomes on the overall instructor ranking.


\section{Conclusion}
We present a preliminary quantitative analysis of sources of potential bias on academic forums. We find that for frequent users of one such academic forum, aggregate student ratings of course instructors gravitate, in many cases disproportionately, towards course outcomes and away from student perception of the relative ease of course materials, content and evaluations. We intend to generalize this analysis into a robust approach of isolating and correcting for bias on academic forums.

\bibliographystyle{ACM-Reference-Format}
\bibliography{sample-bibliography}


\begin{thebibliography}{15}


\ifx \showCODEN    \undefined \def \showCODEN     #1{\unskip}     \fi
\ifx \showDOI      \undefined \def \showDOI       #1{#1}\fi
\ifx \showISBNx    \undefined \def \showISBNx     #1{\unskip}     \fi
\ifx \showISBNxiii \undefined \def \showISBNxiii  #1{\unskip}     \fi
\ifx \showISSN     \undefined \def \showISSN      #1{\unskip}     \fi
\ifx \showLCCN     \undefined \def \showLCCN      #1{\unskip}     \fi
\ifx \shownote     \undefined \def \shownote      #1{#1}          \fi
\ifx \showarticletitle \undefined \def \showarticletitle #1{#1}   \fi
\ifx \showURL      \undefined \def \showURL       {\relax}        \fi
\providecommand\bibfield[2]{#2}
\providecommand\bibinfo[2]{#2}
\providecommand\natexlab[1]{#1}
\providecommand\showeprint[2][]{arXiv:#2}

\bibitem[\protect\citeauthoryear{??}{rmp}{2019}]%
        {rmphomepage}
 \bibinfo{year}{2019}\natexlab{}.
\newblock \bibinfo{title}{Rate My Professors - review teachers and professors,
  school reviews, college campus ratings}.
\newblock   (\bibinfo{year}{2019}).
\newblock
\urldef\tempurl%
\url{http://ratemyprofessors.com/}
\showURL{%
Retrieved 2019 from \tempurl}


\bibitem[\protect\citeauthoryear{Brown, Baillie, and Fraser}{Brown
  et~al\mbox{.}}{2009}]%
        {brown2009rating}
\bibfield{author}{\bibinfo{person}{Michael~J Brown}, \bibinfo{person}{Michelle
  Baillie}, {and} \bibinfo{person}{Shawndel Fraser}.}
  \bibinfo{year}{2009}\natexlab{}.
\newblock \showarticletitle{Rating RateMyProfessors. com: A comparison of
  online and official student evaluations of teaching}.
\newblock \bibinfo{journal}{\emph{College Teaching}} \bibinfo{volume}{57},
  \bibinfo{number}{2} (\bibinfo{year}{2009}), \bibinfo{pages}{89--92}.
\newblock


\bibitem[\protect\citeauthoryear{Centra}{Centra}{2003}]%
        {centragradestudy}
\bibfield{author}{\bibinfo{person}{John~A Centra}.}
  \bibinfo{year}{2003}\natexlab{}.
\newblock \showarticletitle{Will teachers receive higher student evaluations by
  giving higher grades and less course work?}
\newblock \bibinfo{journal}{\emph{Research in Higher Education}}
  \bibinfo{volume}{44}, \bibinfo{number}{5} (\bibinfo{year}{2003}),
  \bibinfo{pages}{495--518}.
\newblock


\bibitem[\protect\citeauthoryear{Chang and Park}{Chang and Park}{2014}]%
        {chang2014koofers}
\bibfield{author}{\bibinfo{person}{YunJeong Chang} {and}
  \bibinfo{person}{Seung~Won Park}.} \bibinfo{year}{2014}\natexlab{}.
\newblock \showarticletitle{Exploring Students' Perspectives of College STEM:
  An Analysis of Course Rating Websites.}
\newblock \bibinfo{journal}{\emph{International Journal of Teaching and
  Learning in Higher Education}} \bibinfo{volume}{26}, \bibinfo{number}{1}
  (\bibinfo{year}{2014}), \bibinfo{pages}{90--101}.
\newblock


\bibitem[\protect\citeauthoryear{Feldman}{Feldman}{1989}]%
        {feldman1989association}
\bibfield{author}{\bibinfo{person}{Kenneth~A Feldman}.}
  \bibinfo{year}{1989}\natexlab{}.
\newblock \showarticletitle{The association between student ratings of specific
  instructional dimensions and student achievement: Refining and extending the
  synthesis of data from multisection validity studies}.
\newblock \bibinfo{journal}{\emph{Research in Higher education}}
  \bibinfo{volume}{30}, \bibinfo{number}{6} (\bibinfo{year}{1989}),
  \bibinfo{pages}{583--645}.
\newblock


\bibitem[\protect\citeauthoryear{Feldman}{Feldman}{2007}]%
        {feldman2007review}
\bibfield{author}{\bibinfo{person}{Kenneth~A Feldman}.}
  \bibinfo{year}{2007}\natexlab{}.
\newblock \showarticletitle{Identifying exemplary teachers and teaching:
  Evidence from student ratings}.
\newblock In \bibinfo{booktitle}{\emph{The scholarship of teaching and learning
  in higher education: An evidence-based perspective}}.
  \bibinfo{publisher}{Springer}, \bibinfo{pages}{93--143}.
\newblock


\bibitem[\protect\citeauthoryear{Glynn~LoPresti and Le}{Glynn~LoPresti and
  Le}{2019}]%
        {koofershomepage}
\bibfield{author}{\bibinfo{person}{Dan~Donahoe Glynn~LoPresti, Patrick~Gartlan}
  {and} \bibinfo{person}{Minhe Le}.} \bibinfo{year}{2019}\natexlab{}.
\newblock \bibinfo{title}{Koofers - professor ratings, practice exams and flash
  cards}.
\newblock   (\bibinfo{year}{2019}).
\newblock
\urldef\tempurl%
\url{http://koofers.com/}
\showURL{%
Retrieved 2019 from \tempurl}


\bibitem[\protect\citeauthoryear{Hartman and Hunt}{Hartman and Hunt}{2013}]%
        {hartman2013ratemyprofessors}
\bibfield{author}{\bibinfo{person}{Katherine~B Hartman} {and}
  \bibinfo{person}{James~B Hunt}.} \bibinfo{year}{2013}\natexlab{}.
\newblock \showarticletitle{What RateMyProfessors. com reveals about how and
  why students evaluate their professors: A glimpse into the student mind-set}.
\newblock \bibinfo{journal}{\emph{Marketing Education Review}}
  \bibinfo{volume}{23}, \bibinfo{number}{2} (\bibinfo{year}{2013}),
  \bibinfo{pages}{151--162}.
\newblock


\bibitem[\protect\citeauthoryear{Hassan and McCrickard}{Hassan and
  McCrickard}{2019}]%
        {hassan2019trust}
\bibfield{author}{\bibinfo{person}{Taha Hassan} {and} \bibinfo{person}{D~Scott
  McCrickard}.} \bibinfo{year}{2019}\natexlab{}.
\newblock \showarticletitle{Trust and Trustworthiness in Social Recommender
  Systems}.
\newblock \bibinfo{journal}{\emph{Companion Proceedings of the 2019 World Wide
  Web Conference (WWW `19 Companion)}} (\bibinfo{date}{May}
  \bibinfo{year}{2019}).
\newblock


\bibitem[\protect\citeauthoryear{Kindred and Mohammed}{Kindred and
  Mohammed}{2005}]%
        {kindred2005he}
\bibfield{author}{\bibinfo{person}{Jeannette Kindred} {and}
  \bibinfo{person}{Shaheed~N Mohammed}.} \bibinfo{year}{2005}\natexlab{}.
\newblock \showarticletitle{``He will crush you like an academic ninja!'':
  Exploring teacher ratings on ratemyprofessors. com}.
\newblock \bibinfo{journal}{\emph{Journal of Computer-Mediated Communication}}
  \bibinfo{volume}{10}, \bibinfo{number}{3} (\bibinfo{year}{2005}),
  \bibinfo{pages}{JCMC10314}.
\newblock


\bibitem[\protect\citeauthoryear{Legg and Wilson}{Legg and Wilson}{2012}]%
        {legg2012ratemyprofessors}
\bibfield{author}{\bibinfo{person}{Angela~M Legg} {and}
  \bibinfo{person}{Janie~H Wilson}.} \bibinfo{year}{2012}\natexlab{}.
\newblock \showarticletitle{RateMyProfessors. com offers biased evaluations}.
\newblock \bibinfo{journal}{\emph{Assessment \& Evaluation in Higher
  Education}} \bibinfo{volume}{37}, \bibinfo{number}{1} (\bibinfo{year}{2012}),
  \bibinfo{pages}{89--97}.
\newblock


\bibitem[\protect\citeauthoryear{Marsh}{Marsh}{1984}]%
        {marsh1984}
\bibfield{author}{\bibinfo{person}{Herbert~W Marsh}.}
  \bibinfo{year}{1984}\natexlab{}.
\newblock \showarticletitle{Students' evaluations of university teaching:
  Dimensionality, reliability, validity, potential baises, and utility.}
\newblock \bibinfo{journal}{\emph{Journal of educational psychology}}
  \bibinfo{volume}{76}, \bibinfo{number}{5} (\bibinfo{year}{1984}),
  \bibinfo{pages}{707}.
\newblock


\bibitem[\protect\citeauthoryear{Meila}{Meila}{1999}]%
        {meila1999accelerated}
\bibfield{author}{\bibinfo{person}{Marina Meila}.}
  \bibinfo{year}{1999}\natexlab{}.
\newblock \showarticletitle{An accelerated Chow and Liu algorithm: fitting tree
  distributions to high dimensional sparse data}.
\newblock  (\bibinfo{year}{1999}).
\newblock


\bibitem[\protect\citeauthoryear{Sch{\"o}lkopf, Smola, and
  M{\"u}ller}{Sch{\"o}lkopf et~al\mbox{.}}{1997}]%
        {scholkopf1997kernel}
\bibfield{author}{\bibinfo{person}{Bernhard Sch{\"o}lkopf},
  \bibinfo{person}{Alexander Smola}, {and} \bibinfo{person}{Klaus-Robert
  M{\"u}ller}.} \bibinfo{year}{1997}\natexlab{}.
\newblock \showarticletitle{Kernel principal component analysis}. In
  \bibinfo{booktitle}{\emph{International conference on artificial neural
  networks}}. Springer, \bibinfo{pages}{583--588}.
\newblock


\bibitem[\protect\citeauthoryear{Stumpf and Freedman}{Stumpf and
  Freedman}{1979}]%
        {stumpf_apa}
\bibfield{author}{\bibinfo{person}{Stephen~A Stumpf} {and}
  \bibinfo{person}{Richard~D Freedman}.} \bibinfo{year}{1979}\natexlab{}.
\newblock \showarticletitle{Expected grade covariation with student ratings of
  instruction: Individual versus class effects.}
\newblock \bibinfo{journal}{\emph{Journal of Educational Psychology}}
  \bibinfo{volume}{71}, \bibinfo{number}{3} (\bibinfo{year}{1979}),
  \bibinfo{pages}{293}.
\newblock


\end{thebibliography}

\end{document}